\documentclass[preprint2,twoside]{hwo}

\usepackage{lipsum}

\input{hwo.h}

\begin{document}

\title{\textbf{\LARGE The Pollux European instrument concept for HWO: a high-resolution spectrograph and spectropolarimeter from the far-UV to the near-IR}}

\author {\textbf{\large Coralie Neiner,$^1$ 
Jean-Claude Bouret,$^2$ 
Luca Fossati,$^3$ 
David le Mignant,$^2$ 
Eduard Muslimov,$^{4,2}$ 
Ana In\'es G\'omez de Castro,$^5$
Fr\'ed\'eric Marin$^6$
}}

\affil{$^1$\small\it LIRA, Paris Observatory, CNRS, PSL University, Sorbonne University, Universit\'e Paris Cit\'e, CY Cergy Paris University, 5 place Jules Janssen, 92195 Meudon, France}
\affil{$^2$\small\it Aix Marseille University, CNRS, CNES, LAM, Marseille, France}
\affil{$^3$\small\it Space Research Institute, Austrian Academy of Sciences, Schmiedlstrasse 6, 8042 Graz, Austria}
\affil{$^4$\small\it Department of Physics, University of Oxford,
Keble Rd, OX14 3RH Oxford, UK}
\affil{$^5$\small\it AEGORA Research Group – Joint Center for Ultraviolet Astronomy, Universidad Complutense de Madrid, Plaza de Ciencias 3,
28040 Madrid, Spain}
\affil{$^6$\small\it Universit\'e de Strasbourg, CNRS, Observatoire Astronomique de Strasbourg, UMR 7550, 11 rue de l’universit\'e, 67000 Strasbourg, France}

% This section is for ADS Processing.  There must be one line per author. Leave them commented out for the present. They will be included later.
%\paperauthor{Sample~Author1}{Author1Email@email.edu}{ORCID_Or_Blank}{Author1 Institution}{Author1 Department}{City}{State/Province}{Postal Code}{Country}
%\paperauthor{Sample~Author2}{Author2Email@email.edu}{ORCID_Or_Blank}{Author2 Institution}{Author2 Department}{City}{State/Province}{Postal Code}{Country}
%\paperauthor{Sample~Author3}{Author3Email@email.edu}{ORCID_Or_Blank}{Author3 Institution}{Author3 Department}{City}{State/Province}{Postal Code}{Country}

% Please provide entries for the Author index
%\aindex{Author, F.}
%\aindex{Author, S.}
%\aindex{Author, T.}

\begin{abstract}
Pollux is a high-resolution spectrograph and spectropolarimeter working from 100 nm to 1.8 microns proposed for HWO by a European consortium. Pollux will allow us to study stellar and (exo)planetary systems, as well as cosmic ecosystems. For example, Pollux will provide new insights on exoplanet formation and evolution, characterization of the atmospheres and magnetospheres of stars and planets, and star-planet interactions. It will also allow us to resolve narrow UV emission and absorption lines, enabling us to follow the baryon cycle over cosmic time --  from galaxies forming stars out of interstellar gas and grains, and planets forming in circumstellar disks, to the various forms of feedback into the interstellar and intergalactic medium -- and from active galactic nuclei. The most innovative characteristic of Pollux is its unique spectropolarimetric capability in the UV, which will open a new parameter space. Its very high spectral resolution ($\sim$70000 to $\sim$100000) and stability over a very large wavelength range will also be a major asset. In this paper, we summarize the main scientific drivers of Pollux and present its current design, technological challenges, and the Pollux consortium organization.\\
\end{abstract}

\vspace{2cm}

\section{Introduction}

Pollux is an instrument proposed by a European consortium for the Habitable Worlds Observatory \citep[HWO, see][and this volume]{Dressing2024}. 

HWO is a flagship mission led by NASA, aimed to be launched in the early 2040's. HWO will be a multi-purpose observatory covering all domains of astrophysics, as well as an exo-Earth hunter. HWO is currently in a maturation phase, called GOMaP (Great Observatory Maturation Program), that will lead to the selection of its telescope concept and suite of instruments in $\sim$ 2030 \citep[see][]{Feinberg2024}.  

Pollux is a high-resolution spectrograph, with spectropolarimetric capabilities covering a very wide wavelength band from the FUV to the NIR. It is proposed to complement other instruments currently considered for HWO, i.e. a coronagraph, a high-resolution imager, a UV multi-object  spectrograph (MOS), and an integral field unit (IFU). 

In this paper, we present examples of the science that Pollux will tackle (Sect.~\ref{sect_Pollux_science}), the current preliminary instrument design (Sect.~\ref{sect_Pollux_design}), the required technological developments (Sect.~\ref{sect_Pollux_tech}), and the organization of the Pollux consortium (Sect.~\ref{sect_Pollux_org}). 

\begin{figure*}[ht]
\begin{center}
\includegraphics[width=\textwidth]{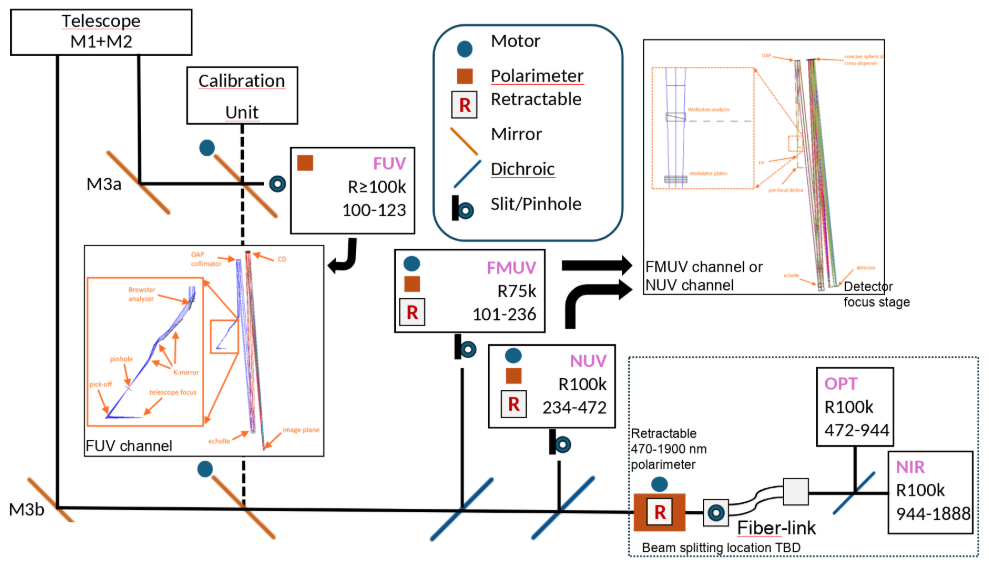}
\caption{\small Schematic view of the current Pollux design. The caption of symbols is shown at the top center of the figure. Light flows along the black lines. The FMUV and NUV channel designs are similar, therefore only one example is shown.   
\label{schema_pollux}
}
\end{center}
\end{figure*}

\section{Pollux Science}
\label{sect_Pollux_science}

The main science cases of Pollux cover similar domains as the main science cases of HWO and have been organized in 3 themes: stars, (exo-)planets, and cosmic ecosystems. Several of them have been submitted as Science Case Development Documents (SCDD) for HWO. 

\subsection{Stars}

The Stars theme includes, for example, the impact magnetic fields may have on the formation of stars and of their planets, including studying the link between magnetic fields and accretion/ejection flows in protostars to test the magnetospheric accretion scenario. It also includes defining new frontiers in the study of magnetic massive stars by determining the origin of magnetism in massive stars by testing several scenarios (fossil field, binary merger,...), determining the mutual roles these stars and their magnetic fields play into each other's evolution, and exploring magnetism in massive stars in environments with different metallicities (see David-Uraz et al., this volume). Pollux will also allow  performing stellar spectroscopy in the Local group and beyond. It will provide comprehensive observational constraints on the evolution of very metal-poor massive stars, detailed chemistry and metallicity distributions for first and second generation stars, and help study cepheids in the Local Group and beyond, e.g. by calibrating the period-luminosity-metallicity relations in various galaxies. 

\subsection{Planets and exoplanets}

The (Exo-)planets theme includes solar system science, such as the study of the impact of energetic particle precipitation (aurorae) on the atmospheres of giant solar system planets and charge exchange in comets. In addition, Pollux will allow the characterization of exoplanet atmospheres and atmospheric escape, via the detection of atomic and molecular species from the UV to the NIR, and the study of the impact of photochemistry on upper and lower atmospheres. The high-resolution high-signal-to-noise spectra will contain no telluric lines, retain information from the continuum light, and record the signature of molecules in the NUV, allowing for exquisite science. It will thus be possible to study a wide range of exoplanets, from habitable-zone rocky planets to close-in giant planets. Pollux will also provide unprecedented information on star-planet interactions, in particular the magnetic connection, tidal interactions, space weather, and exoplanetary magnetic fields. It will also be possible to study sputtered surface material from close-in rocky planets and characterize metals in the atmospheres of DZ white dwarfs. See Fossati et al. (this volume) for more information on the Pollux exoplanet science case. 

\begin{figure*}[ht]
\begin{center}
\includegraphics[width=\textwidth]{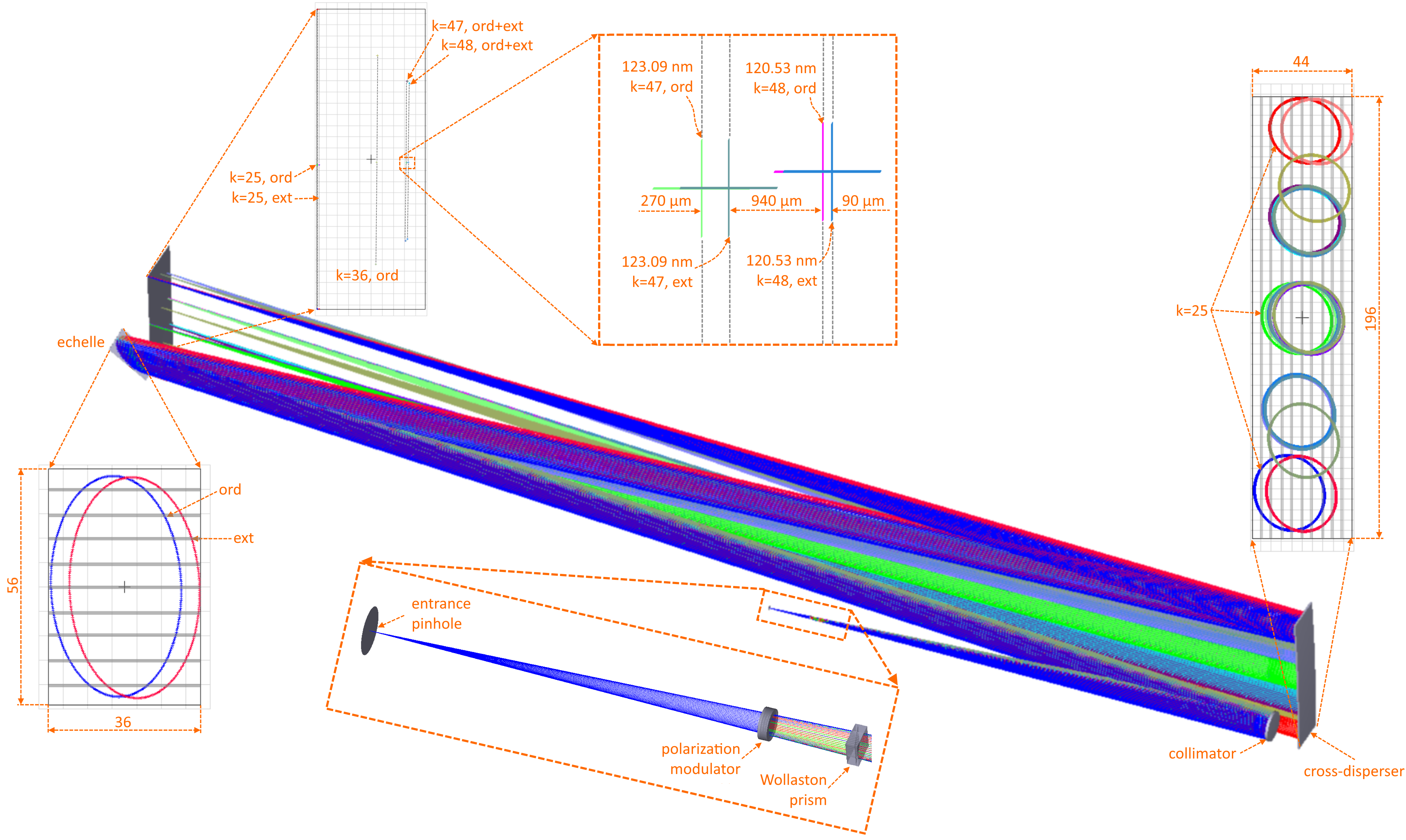}
\caption{\small Optical design of an echelle spectropolarimeter on the example of FMUV channel. 
\label{FMUV_inDetails}
}
\end{center}
\end{figure*}

\subsection{Cosmic ecosystems}

The Cosmic Ecosystems theme regroups the Pollux science cases related to interstellar (ISM), circumgalactic (CGM), and intergalactic (IGM) media, quiescent, active, and interacting galaxies, and local and cosmological probes. For example, with Pollux it will be possible to measure the polarization by diffusion and absorption in gas and dust in the ISM, CGM, and IGM. We will understand the role of magnetic fields in these media and the structuring of the media, as well as interactions between galaxies and their environment and dust chemistry. We will be able to study the dynamics of galactic disks and halos, the physics of accretion and general relativity, feedback from AGNs (jets, winds) and starbursts, and the impact of the interaction between galaxies on their evolution. Pollux will also measure traces of reionization and first objects, the polarization of fossil radiation and distant galaxies, provide constraints on dark matter and dark energy, and observe cosmic gravitational lensing. Also see Marin et al. and Biedermann et al. (this volume) about the study of  massive black holes with Pollux.\\

Therefore, the Pollux scientific goals cover a wide range of astrophysics and align very well with the scientific priorities of the NASA Astrophysics Decadal survey 2020, the ESA Voyage 2050, and the Astronet roadmap. 

Recently, the Pollux project office issued a call for additional science cases, with the goal of refining the instrumental design and maximizing the mission’s scientific return. This will help converge towards the best instrument for the whole community. 

\section{Instrument design}
\label{sect_Pollux_design}

To reach the scientific goals presented in Sect.~\ref{sect_Pollux_science}, we have designed a preliminary version of Pollux. A schematic view of the Pollux design is shown in Fig~\ref{schema_pollux}. It is composed of 5 high-resolution echelle spectrographs, each equipped with a dedicated polarimeter, covering from the FUV to the NIR. The current 5 channels are defined as follows:
\begin{itemize}
    \item FUV: 100-123 nm, R $\geq$ 90 000, working in spectropolarimetric mode only. 
    \item FMUV: 101-236 nm, R $\geq$ 71 000, with a retractable polarimeter for the 120.5-236 nm range and no polarimetry in the 101-120.5 nm range. 
    \item NUV: 234-472 nm, R $\geq$ 94 000, with a retractable polarimeter for the full range.
    \item OPT: 472-944 nm, R $\sim$ 100 000, with a retractable polarimeter for the full range.
    \item NIR: 944-1888 nm, R $\sim$ 100 000, with a retractable polarimeter for the full range.
\end{itemize}

\begin{figure*}[ht]
\begin{center}
\includegraphics[width=0.8\textwidth]{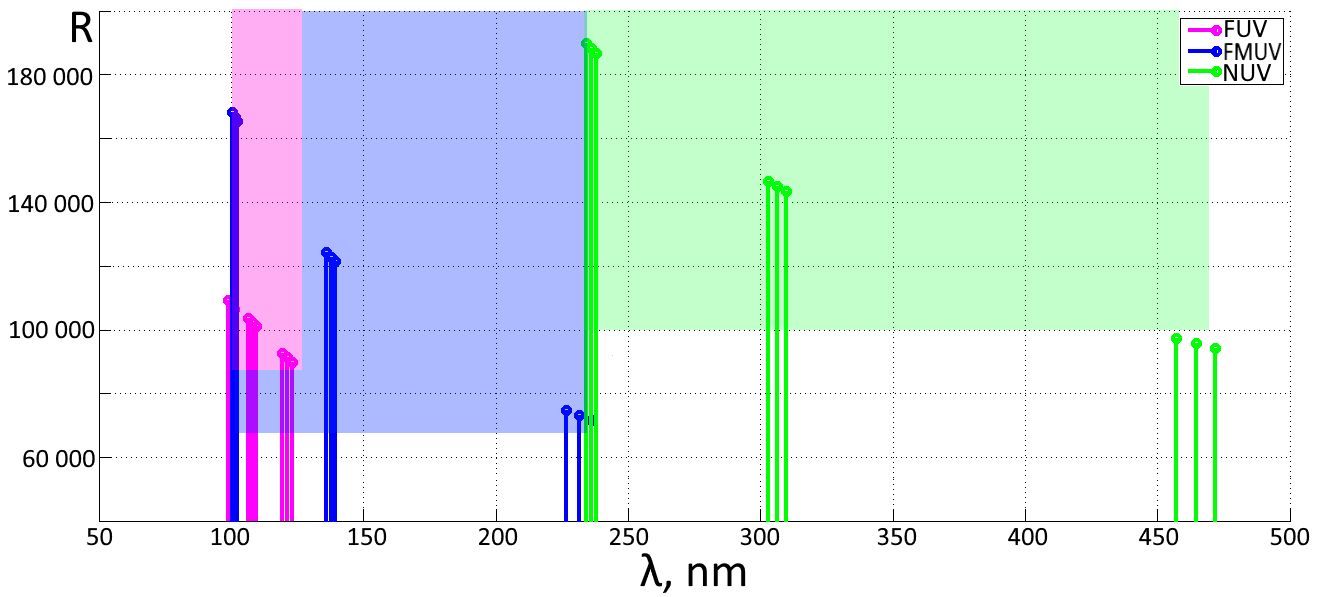}
\caption{\small Spectral resolving power for reference wavelengths of the UV channels. The shaded zones correspond to the target values.  
\label{UV_resolving_power}
}
\end{center}
\end{figure*}

The polarimeters measure both circular and linear polarization, i.e. all four Stokes parameters (I, Q, U, V). The FUV polarization modulator consists of mirrors at grazing incidence and the analyzer is a single mirror at the Brewster angle, therefore only one beam is recorded at the exit of the analyzer. In addition, such polarimeters made of mirrors are very sensitive to alignment, therefore the FUV polarimeter module cannot be retracted. The other 4 polarimeters are made of thin birefringent plates and Wollaston prisms, which result in recording two beams of orthogonal polarization states after the analyzer. These polarimeters can be retracted to allow for pure spectroscopy mode. Pure FUV spectroscopy is performed thanks to the FMUV channel. 

The echelle spectrograph designs are very similar for each of the 5 channels. To illustrate it, a schematic view of the FMUV channel is shown in Fig.~\ref{FMUV_inDetails}. It shows the ordinary (\textit{ord}) and extraordinary (\textit{ext}) rays at the central and marginal wavelengths of 4 diffraction orders of the echelle ($k=25..48$) traced through the system. The collimator represents an off-axis parabolic mirror, which delivers a collimated beam to the echelle grating. The overlapping orders diffracted by the echelle are separated by the cross-disperser, which also acts as a camera. Thus, the number of surfaces has been minimized to maximize throughput. 

The FUV, FMUV, NUV, and OPT channels use large CMOS detectors, while the NIR channel uses a large H2RG detector, similar to JWST. Micro-channel plates are currently not considered for Pollux, in order to keep the global architecture as simple as possible. In addition, all detectors operate without active cooling.

Pollux can thus be used for point source spectroscopy, covering the full wavelength range from 101 nm to 1.88 $\mu$m in a single exposure. It can also be used for point source spectropolarimetry, either in the FUV (100-123 nm) domain or in the wavelength range from 120.5 nm to 1.88 $\mu$m, i.e. covering the full range from FUV to NIR in 2 measurements. Each spectropolarimetric measurement is composed of 6 sub-exposures at 6 different positions of the modulator. Finally, a slit spectroscopy mode can be used in the FMUV and NUV channels. 

\begin{figure*}[ht]
\begin{center}
\includegraphics[width=\textwidth]{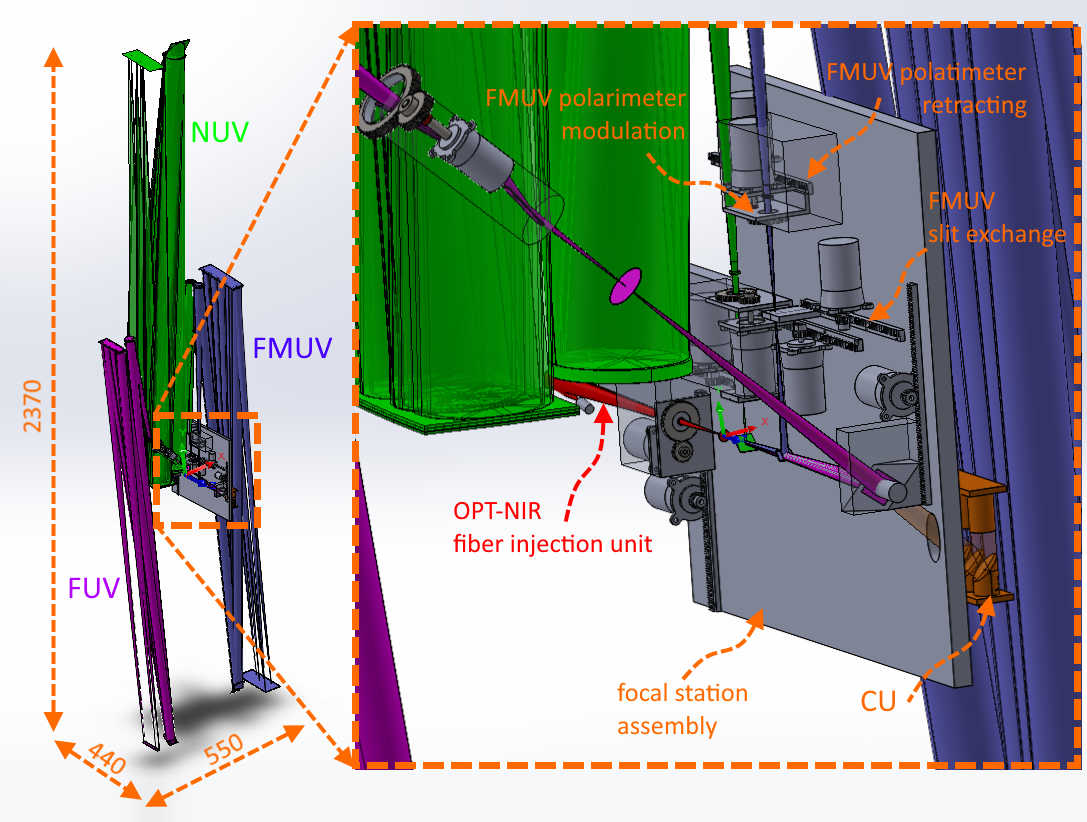}
\caption{\small Packaging and overall dimensions of the three UV sub-systems. 
\label{UV_packaging}
}
\end{center}
\end{figure*}

The spectral resolving power of all channels is very high and limited mainly by the telescope point spread function (PSF) width. The resolving power of the three UV channels for typical wavelengths is shown in Fig.~\ref{UV_resolving_power}. These values are calculated with a presumption that the entrance pinhole of each channel corresponds to the first Airy ring diameter, and the actual image quality will degrade by $\approx 10\%$ with respect to its theoretical value. 
Fig.~\ref{UV_packaging} shows these three UV channels packaged and the focal station that feeds them either with the science beam, or with a beam formed by the on-board calibration unit (CU). In this configuration, the OPT and NIR channels are coupled via a fiber connection. This figure also illustrates a possible implementation of mechanisms for switching between the observation modes described above.    

\begin{figure*}[ht]
\begin{center}
\includegraphics[width=0.8\textwidth]{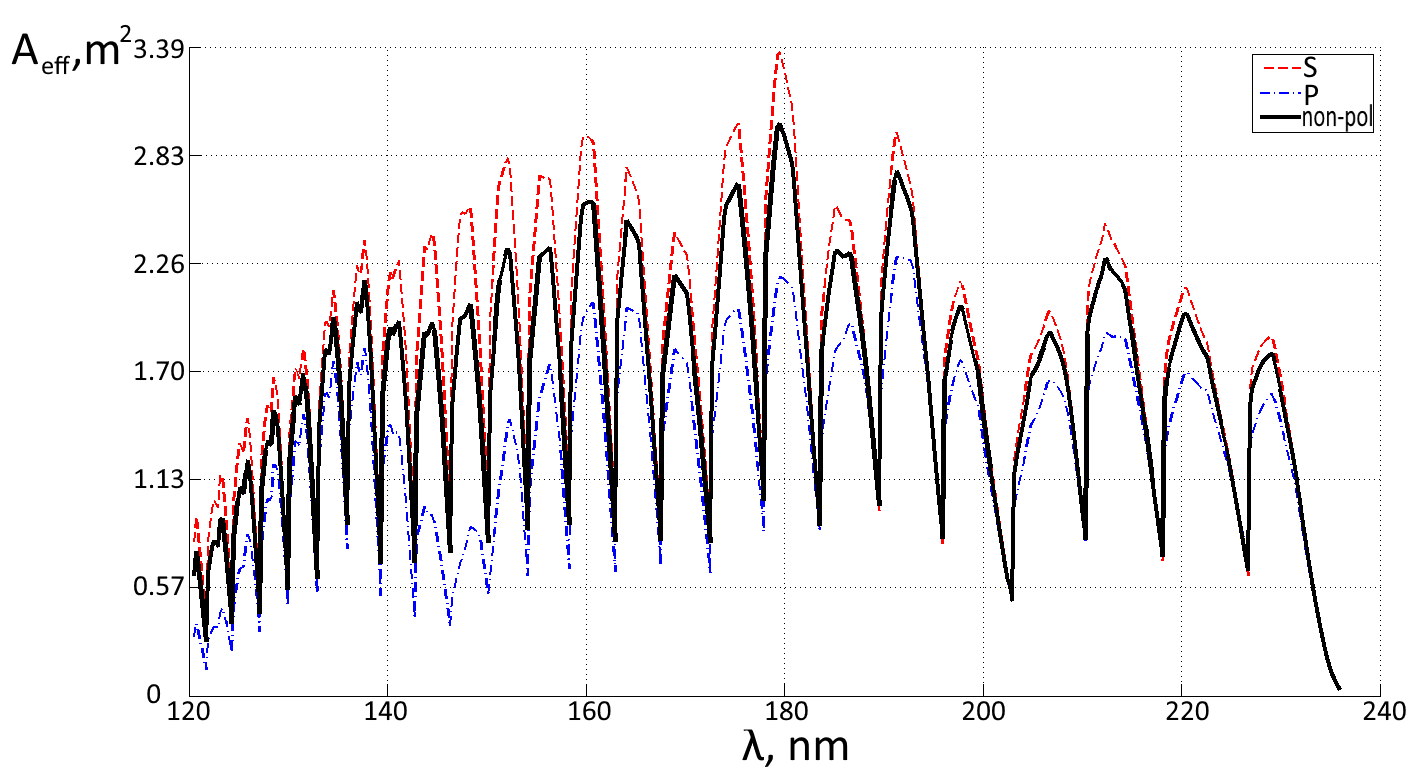}
\caption{\small End-to-end transmission of the FMUV channel in the MUV spectropolarimetric mode presented in the effective area units for a 6-m unobscured telescope. 
\label{MUV_transmission}
}
\end{center}
\end{figure*}

As an example, the current transmission of the FMUV channel in the spectropolarimetric mode is shown as an example in Fig.~\ref{MUV_transmission}. To simplify comparison with other instruments, it is expressed in the units of effective collecting area ($A_{eff}$) for a case of 6-m unobscured telescope. 

As an alternative configuration, we are investigating the possibility to feed the FUV channel directly via a dedicated pick-off mirror (M3) after the primary (M1) and secondary (M2) HWO telescope mirrors. This approach would help to reduce reflections and thus maximize throughput. The other four channels (FMUV, NUV, OPT,  and NIR) would then be fed through a different M3 than the FUV channel, and operate simultaneously thanks to a series of three dichroics that split the light into the four spectrographs. 

The current Pollux design remains preliminary and will evolve in the coming years, but the main requirements are set, namely high-resolution spectroscopy and spectropolarimetry on a wide wavelength coverage.

\section{Technological developments}
\label{sect_Pollux_tech}

Building an instrument as the one proposed above requires developing several technologies. These developments are ongoing in Europe. 

First, in France, polarimeters working in the NUV and MUV wavelength ranges have been developed and tested using MgF2 birefringent plates and Wollaston prisms \citep[see e.g.][]{Pertenais}. This technology is now well advanced \citep{Neiner2025}, but final tests and space qualification still need to be performed. In the meantime, FUV polarimeters using mirrors are under development \citep[see][and Girardot et al., this volume]{Girardot2024} but currently have a much lower TRL.

Dichroics separating the UV and Visible range have also been studied in France through a collaboration with REOSC. The three dichroics of Pollux must be very efficient and should work on very wide wavelength bands. In particular, the first dichroic that must separate the FMUV range (101-236 nm) from the longer wavelengths (up to 1.88 $\mu$m) is very challenging.

The gratings used in the spectrographs, both echelles and cross-dispersers, must also be efficient and should not produce too much polarization. Indeed, while each polarimeter is placed before its corresponding spectrograph in the light path, and thus the polarimetric information is already encoded in the intensity of the beams, it is still important to keep a good balance of the flux level between two orthogonal polarization states. This would otherwise decrease the signal-to-noise and thus precision of the final polarimetric measurement. Development work for large, non-polarizing, UV gratings is currently being set up with Horiba. The echelle gratings are expected to be engraved, while the cross-dispersers are likely to be holographic. 

The CMOS used for the UV channels must be efficient, which requires new coatings. Studies are underway in the UK (see Skottfeld et al., this volume), Spain, as well as in the USA, in the context of HWO and in collaboration with Teledyne.

Finally, to get rid of UV calibration lamps, which have a short lifetime and would require a lot of redundancy, we are investigating the possibility of using a Fabry-Perot comb in the UV domain. This development work will be performed in collaboration between Belgium, the Czech Republic, Austria, and Germany. No UV comb exists as of today. 

\section{Pollux organization}
\label{sect_Pollux_org}

\subsection{Pollux consortium}

\begin{figure*}[ht]
\begin{center}
\includegraphics[width=\textwidth]{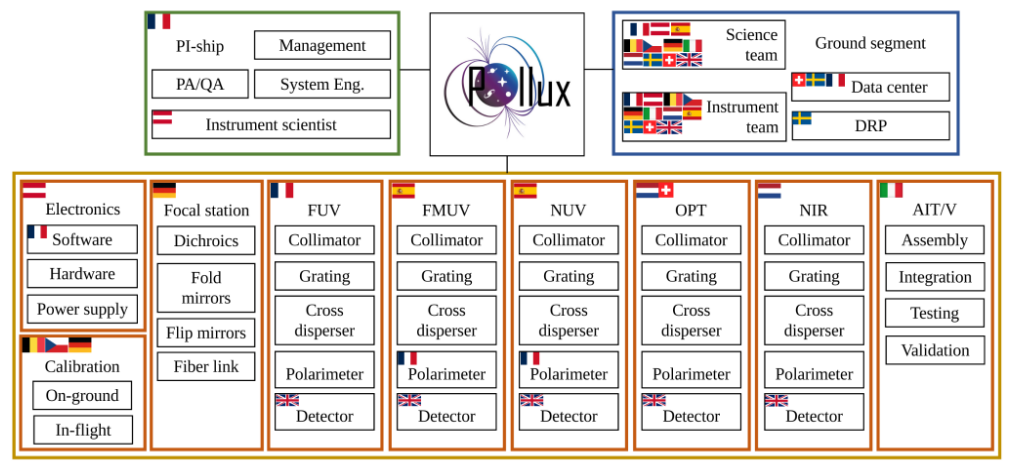}
\caption{\small Currently projected European consortium for Pollux. National contributions are subject to the formal approval of the respective agencies. 
\label{pollux_consortium}
}
\end{center}
\end{figure*}

To prepare and build Pollux, a European consortium has been formed, which includes 11 countries and is led by France. The various contributions currently considered for each country are summarized in Fig.~\ref{pollux_consortium} and listed below. However, these national contributions are still subject to the formal approval of the respective agencies and could evolve in the coming years. 
\begin{itemize}
    \item France is the PI country, in charge of management, system engineering, product and quality assurance, of the FUV spectrograph including its polarimeter but excluding its detection chain, of the polarimeters of the FMUV and NUV channels, of the electronics software, and will contribute to the data center;
    \item Austria has the instrument scientist role and is in charge of the electronics hardware and power supply;
    \item Belgium is in charge of both on-ground and in-flight calibration; 
    \item the Czech Republic will contribute to the calibration;
    \item Germany is in charge of the focal station, which includes the dichroics, fold and flip mirrors, pinholes and slits, and fiber link. It requires several motors. See and example for the UV channels in Fig.~\ref{UV_packaging}; 
    \item Italy is currently in charge of the global AIT/AIV of the instrument, but is also considering taking responsibility for one spectrograph channel; 
    \item Spain is in charge of the FMUV and NUV spectrographs, except their polarimeters; 
    \item the Netherlands are in charge of the OPT and NIR spectrographs and their polarimeters; 
    \item Sweden is in charge of the data reduction pipeline in the ground segment and will contribute to the data center; 
    \item Switzerland will contribute to the OPT spectrograph and to the data center; 
    \item UK is in charge of all detection chains.
\end{itemize}

In addition, all 11 countries participate in the Pollux instrument team and science team. A Pollux project office is formed of the two principal investigators (PI, Coralie Neiner and Jean-Claude Bouret), the project manager (David Le Mignant), and the instrument scientist (Luca Fossati). 

\subsection{Pollux funding}

The Pollux instrument can be funded in several ways. First, the Pollux consortium composed of the 11 European countries can fund Pollux on their national space agency or research budgets. Alternatively, ESA is expected to issue a call for contributions to HWO in the coming years, to which Pollux can be submitted by the consortium to obtain ESA funding. This funding can cover part or all of the budget needed for Pollux and can be complemented by the national budgets of the 11 countries.\\

%\acknowledgements
{\bf The Pollux consortium acknowledges support for technological developments from the French space agency CNES.}

\bibliography{author.bib}

\end{document}